# Normal solutions of the Boltzmann equation for highly nonequilibrium

# Fourier flow and Couette flow

M. A. Gallis, J. R. Torczynski, and D. J. Rader

*Engineering Sciences Center, Sandia National Laboratories, Albuquerque, New Mexico 87185-0826 USA*

M. Tij

*Département de Physique, Université Moulay Ismaïl, Meknès, Morocco*

A. Santos

*Departamento de Física, Universidad de Extremadura, E-06071 Badajoz, Spain*

Corresponding Author:

John R. Torczynski, 505-845-8991 (work), 505-844-8251 (fax), jrtorcz@sandia.gov

## ABSTRACT

The state of a single-species monatomic gas from near-equilibrium to highly nonequilibrium conditions is investigated using analytical and numerical methods. Normal solutions of the Boltzmann equation for Fourier flow (uniform heat flux) and Couette flow (uniform shear stress) are found in terms of the heat-flux and shear-stress Knudsen numbers. Analytical solutions are found for inverse-power-law molecules from hard-sphere through Maxwell at small Knudsen numbers using Chapman-Enskog (CE) theory and for Maxwell molecules at finite Knudsen numbers using a moment-hierarchy (MH) method. Corresponding numerical solutions are obtained using the Direct Simulation Monte Carlo (DSMC) method of Bird. The thermal conductivity, the viscosity, and the Sonine-polynomial coefficients of the velocity distribution function from DSMC agree with CE results at small Knudsen numbers and with MH results at finite Knudsen numbers. Subtle differences between inverse-power-law, variable-soft-sphere, and variable-hard-sphere representations of Maxwell molecules are observed. The MH and DSMC results both indicate that the effective thermal conductivity and the effective viscosity for Maxwell molecules are independent of the heat-flux Knudsen number, and additional DSMC simulations indicate that these transport properties for hard-sphere molecules decrease slightly as the heat-flux Knudsen number is increased. Similarly, the MH and DSMC results indicate that the effective thermal conductivity and the effective viscosity for Maxwell molecules decrease as the shear-stress Knudsen number is increased, and additional DSMC simulations indicate the same behavior for hard-sphere molecules. These results provide strong evidence that the DSMC method can be used to determine the state of a gas under highly nonequilibrium conditions.

## I. INTRODUCTION

The state of a single-species monatomic gas under highly nonequilibrium conditions remains a fundamental research problem with important applications. Departure from equilibrium can be achieved in two ways: through rarefaction effects (molecules collide with solid boundaries more frequently than with each other) and through gradients in flow properties. Herein, the latter situation is considered: the heat flux or the shear stress is finite rather than infinitesimal, whereas the direct influence of solid boundaries is negligible. This situation can exist in heated micron-scale devices surrounded by air at ambient conditions.

The Boltzmann equation (BE) is the starting point for kinetics-based investigations into the nonequilibrium properties of gases.[1] At the level of the BE, a "hydrodynamic" description is obtained when the velocity distribution function is a "normal" solution: temporal and spatial variations occur entirely through a functional dependence on the hydrodynamic fields (the number density, the temperature, and the velocity). An equivalent way of posing this is to consider two types of Knudsen numbers: a *system* Knudsen number, defined as the ratio of the mean free path to a characteristic geometric length scale, and a *local* Knudsen number, defined as the ratio of the mean free path to a local characteristic hydrodynamic length scale determined from the local heat flux or shear stress. If both types of Knudsen numbers are small, then the flow is hydrodynamic and continuum, and the Navier-Stokes equations apply. If the system Knudsen number is not small, then the flow is not hydrodynamic, with noncontinuum effects produced by molecule collisions with solid surfaces. On the other hand, if the system Knudsen number is small but the local Knudsen number is finite, then the flow is hydrodynamic, with noncontinuum effects produced by flow gradients. In this latter situation, a normal solution of the BE can be obtained.

Despite the physical simplicity of the BE, its mathematical complexity makes it difficult to find exact solutions. The main source of this complexity is the collision term, which is an integral over velocity space of quadratic terms involving the velocity distribution function. In hydrodynamic continuum situations, the departure from equilibrium is small, so the velocity distribution function can be represented as a perturbation expansion about the equilibrium distribution in terms of the local Knudsen number. Chapman-Enskog (CE) theory[1] provides such a representation of the velocity distribution function. The first-order CE solution, however, is limited to small local Knudsen numbers (i.e., near-equilibrium continuum conditions).

The collision term becomes more tractable for certain molecular interactions. In the case of Maxwell molecules, which repel each other with a force inversely proportional to the fifth power of the distance between their centers, moments of the collision term can be determined directly from moments of the velocity distribution function without requiring detailed knowledge of this function. This moment property for Maxwell molecules allows a hierarchy of moment equations to be derived from the BE, and this system of equations admits the possibility of recursively obtaining normal solutions. Several exact solutions for uniform heat flux and uniform shear stress (Fourier flow and Couette flow) that are based on this moment-hierarchy (MH) method have appeared in the literature,[2-7] and efficient algorithms for computing these moments symbolically and numerically have been reported.[8] Corresponding solutions for the Bhatnagar-Gross-Krook (BGK) collision term have also been derived.[9,10]

The Direct Simulation Monte Carlo (DSMC) method of Bird[11] is a numerical method for simulating nonequilibrium gas behavior that is based on kinetic theory. In brief, computational molecules move, reflect from solid boundaries, and collide with each other so as to statistically mimic the behavior of real molecules. Wagner[12] provides a rigorous proof that DSMC produces a solution to the Boltzmann equation in the limit of vanishing discretization and stochastic errors. Santos and co-workers apply DSMC to Fourier flow and Couette flow with Maxwell and hard-sphere molecules and use the results to evaluate the accuracy of BGK-like and Grad models.[13,14] Gallis and co-workers[15-17] show that DSMC reproduces the infinite-approximation CE results for the thermal conductivity and the viscosity of Maxwell and hard-sphere molecules.

Herein, two classical benchmark problems are investigated in which the flow properties vary one-dimensionally in space and remain constant in time. In Fourier flow, the gas experiences a uniform heat flux, whereas in Couette flow, the gas experiences a uniform shear stress. These flows are produced in the DSMC simulations by confining the gas between two parallel, solid walls at fixed, uniform conditions. To ensure that a normal solution is obtained in the central region of the domain, the walls are separated by about 40 mean free paths.

Fourier flow, shown in Figure 1, is perhaps the simplest situation for studying gas behavior under highly nonequilibrium conditions. Gas is confined between two infinite, parallel walls separated by a distance $L$ that are motionless (tangential velocities $V_1 = V_2 = 0$) but have unequal temperatures ($T_1 \neq T_2$). Starting from an arbitrary initial state, the system reaches a steady state after a transient period during which the molecules travel between the walls several times. After steady state is reached, the gas is motionless, and a uniform heat flux and a corresponding temperature gradient exist in the domain. When the heat flux (or, more precisely, the heat-flux Knudsen number discussed below) is small, the heat flux is proportional to the temperature gradient in the bulk gas (i.e., several mean free paths away from the walls) according to Fourier's law, where the coefficient of proportionality is the thermal conductivity.

Couette flow, also shown in Figure 1, is another simple situation for studying gas behavior under highly nonequilibrium conditions. In this situation, the walls are isothermal (temperatures $T_1 = T_2$) but have unequal tangential velocities ($V_1 \neq V_2$). Steady flow is achieved in the same manner as in the previous situation. After steady state is reached, the normal velocity component is zero, and a uniform shear stress and a corresponding tangential-velocity gradient exist in the domain. When the shear stress (or, more precisely the shear-stress Knudsen number discussed below) is small, the shear stress is proportional to the tangential-velocity gradient in the bulk gas according to Newton's law, where the coefficient of proportionality is the viscosity. Couette flow is more complicated than Fourier flow because the viscous heat generation is nonzero throughout the domain and therefore the temperature distribution is nonuniform. Since $T_1 = T_2$, the temperature distribution reaches a maximum at the center of the domain. When the walls have unequal temperatures and tangential velocities, both Fourier flow and Couette flow are obtained. More specifically, at small heat fluxes and shear stresses, Fourier's law and Newton's law are jointly observed in the bulk gas well away from the walls.

In the following sections, Fourier flow and Couette flow from near-equilibrium to highly nonequilibrium conditions are studied in detail. Inverse-power-law (IPL) molecular interactions from hard-sphere through Maxwell are examined. At near-equilibrium conditions, analytical solutions from CE theory and numerical solutions from DSMC are compared. For Maxwell molecules at highly nonequilibrium conditions, analytical solutions from the MH method and numerical solutions from DSMC are compared. For hard-sphere molecules at these conditions, only DSMC numerical solutions are presented since no analytical results are available. In all situations, the thermal conductivity, the viscosity, and the Sonine-polynomial coefficients that characterize the velocity distribution function are of particular interest, as is the dependence of each of these quantities on the heat-flux and shear-stress Knudsen numbers.

## II. CHAPMAN-ENSKOG THEORY

## A. General results

Chapman-Enskog (CE) theory provides a method for obtaining the normal solution of the BE in terms of an expansion in the gradients of hydrodynamic flow properties or, equivalently, powers of the heat-flux and shear-stress Knudsen numbers.[1] If truncated at first order in these Knudsen numbers, CE theory describes the state of a nonequilibrium gas in the hydrodynamic limit for a small heat-flux vector and a small shear-stress tensor (i.e., the Navier-Stokes equations). In this situation, first-order CE theory generates a closed-form expression for the velocity distribution function in terms of macroscopic hydrodynamic fields and their gradients:

$$f = f^{(0)}(1 + \Phi^{(1)} + \Psi^{(1)}), \tag{1}$$

$$f^{(0)} = n \exp[-\tilde{c}^2] / \left(\pi^{3/2} c_m^3\right), \tag{2}$$

$$\Phi^{(1)} = -(8/5)\,\tilde{A}[\tilde{c}]\tilde{\mathbf{c}} \cdot \tilde{\mathbf{q}}, \tag{3}$$

$$\Psi^{(1)} = -2\tilde{B}[\tilde{c}](\tilde{\mathbf{c}} \circ \tilde{\mathbf{c}} : \tilde{\tau}). \tag{4}$$

Here, $f^{(0)}$ is the equilibrium (Maxwellian) distribution, $\Phi^{(1)}$ and $\Psi^{(1)}$ are the first-order nonequilibrium perturbations from this distribution, $c_m = \sqrt{2k_B T/m}$ is the most probable molecular thermal speed for the equilibrium distribution, $m$ is the molecular mass, $n$ is the number density, $T$ is the temperature, $k_B$ is the Boltzmann constant, $\mathbf{c} = \mathbf{u} - \mathbf{U}$ is the thermal velocity of a molecule, $\mathbf{u} = (u, v, w)$ is the velocity of a molecule, $\mathbf{U} = (U, V, W) = \langle \mathbf{u} \rangle$ is the average value of $\mathbf{u}$, $\tilde{\mathbf{c}} = \mathbf{c}/c_m$ is the normalized molecular thermal velocity, $\tilde{\mathbf{c}} \circ \tilde{\mathbf{c}} = \tilde{\mathbf{c}}\tilde{\mathbf{c}} - (\tilde{c}^2/3)\mathbf{I}$ is a traceless dyadic, $\tilde{\mathbf{q}} = \mathbf{q}/(mnc_m^3)$ and $\tilde{\tau} = \tau/(mnc_m^2)$ are the nondimensional heat-flux vector and shear-stress tensor, and $\tilde{A}$ and $\tilde{B}$ are expansions in the Sonine polynomials $S_j^{(k)}$:

$$\tilde{A}[\tilde{c}] = \sum_{k=1}^{\infty} (a_k/a_1) S_{3/2}^{(k)} [\tilde{c}^2],$$ (5)

$$\tilde{B}[\tilde{c}] = \sum_{k=1}^{\infty} (b_k/b_1) S_{5/2}^{(k-1)} [\tilde{c}^2],$$ (6)

$$S_j^{(k)}[\xi] = \sum_{i=0}^{k} \frac{(j+k)!(-\xi)^i}{(j+i)!i!(k-i)!}.$$ (7)

The heat-flux vector and the shear-stress tensor obey the constitutive equations below:

$$\mathbf{q} = -K\nabla T,$$ (8)

$$\tau = \mu\left\{ (\nabla\mathbf{U} + \nabla\mathbf{U}^T) - \tfrac{2}{3}(\nabla\cdot\mathbf{U})\mathbf{I} \right\}.$$ (9)

The thermal conductivity $K$ and the viscosity $\mu$ in the above obey the following relations:

$$K = -(5/4)k_B c_m^2 a_1,$$ (10)

$$\mu = (1/2)mc_m^2 b_1,$$ (11)

$$K = \left(\frac{K_\infty}{K_1}\right)\left(\frac{15}{4}\right)\left(\frac{k_B}{m}\right)\left(\frac{\mu_1}{\mu_\infty}\right)\mu.$$ (12)

Here, $K_\infty/K_1$ and $\mu_\infty/\mu_1$ are the CE infinite-to-first-approximation ratios of the thermal conductivity and the viscosity, respectively. The $a_k$ and the $b_k$ are the heat-flux and shear-stress Sonine-polynomial coefficients, respectively. Ratios of these coefficients can be expressed in terms of moments of the velocity distribution function, where, for convenience, the relevant nonzero components of the heat-flux vector and the shear-stress tensor are taken to be $q_x$ and $\tau_{xy}$, respectively:[17]

$$\frac{a_k}{a_1} = \sum_{i=1}^{k} \left( \frac{(-1)^{i-1}k!(5/2)!}{(k-i)!i!(i+3/2)!} \right) \frac{\langle \tilde{c}^{2i}\tilde{c}_x \rangle}{\langle \tilde{c}^2\tilde{c}_x \rangle}$$ (13)

$$\frac{b_k}{b_1} = \sum_{i=1}^{k} \left( \frac{(-1)^{i-1}(k-1)!(5/2)!}{(k-i)!(i-1)!(i+3/2)!} \right) \frac{\left\langle \tilde{c}^{2(i-1)} \tilde{c}_x \tilde{c}_y \right\rangle}{\left\langle \tilde{c}_x \tilde{c}_y \right\rangle} \tag{14}$$

The (local) heat-flux and shear-stress Knudsen numbers are defined to be the nondimensional heat flux and shear stress, respectively:

$$\text{Kn}_q = \left| \tilde{q}_x \right| = \left| q_x \right| / (mnc_m^3), \tag{15}$$

$$\text{Kn}_\tau = \left| \tilde{\tau}_{xy} \right| = \left| \tau_{xy} \right| / (mnc_m^2). \tag{16}$$

## B. Molecular interactions

A molecular interaction must be specified in order to use CE theory to determine the thermal conductivity, the viscosity, and the Sonine-polynomial coefficients. In the inverse-power-law (IPL) interaction, the repulsive force between two molecules varies according to $1/r^\nu$, where $r$ is the distance between their centers and $5 \leq \nu \leq \infty$. The two limiting cases are the Maxwell interaction, for which $\nu = 5$, and the hard-sphere interaction, for which $\nu \to \infty$. The IPL interaction yields a thermal conductivity and a viscosity with the following forms:[1]

$$K = K_{\text{ref}}(T/T_{\text{ref}})^\omega, \tag{17}$$

$$\mu = \mu_{\text{ref}}(T/T_{\text{ref}})^\omega, \tag{18}$$

$$\omega = (1/2) + [2/(\nu-1)], \tag{19}$$

where "ref" denotes reference quantities. The parameter $\omega$ takes the values of 1/2 and 1 for the hard-sphere and Maxwell interactions, respectively.

In DSMC simulations, the variable-soft-sphere (VSS) interaction of Koura and Matsumoto[18] and the variable-hard-sphere (VHS) interaction of Bird[11] are used to approximate the IPL interaction. All three interactions yield thermal conductivities and viscosities with the same temperature dependence, namely proportional to $T^\omega$. The VSS and VHS interactions use a molecular diameter $d$ that depends on the relative molecular speed $c_r$ according to $d \propto c_r^{-2/(\nu-1)}$, where the reference molecular diameter is given by the following expression:[11,15]

$$d_{\text{ref}} = \left( \frac{5(\alpha+1)(\alpha+2)(mk_B T_{\text{ref}}/\pi)^{1/2}}{4\alpha(5-2\omega)(7-2\omega)\mu_{\text{ref}}(\mu_1/\mu_\infty)} \right)^{1/2}. \tag{20}$$

Here, $\alpha$ is the angular-scattering parameter, which relates the scattering angle $\chi$ to the impact parameter $b$ according to $b \propto \cos^\alpha[\chi/2]$. Since $\alpha=1$ produces isotropic (hard-sphere) scattering, the VHS interaction uses this value for all $\omega$ values (i.e., for all IPL interactions).

The VSS interaction uses different values of $\alpha$ to represent different IPL interactions. The $\alpha$ value that achieves the best match between the VSS and IPL interactions is determined by equating the VSS and IPL Schmidt numbers (the Schmidt number is $\mu/\rho D$, where $\mu$ is the viscosity, $\rho = mn$ is the mass density, and $D$ is the self-diffusion coefficient):

$$\alpha = (2A_2[\nu])/(2A_1[\nu] - A_2[\nu]), \tag{21}$$

where $A_1[\nu]$ and $A_2[\nu]$ are functions in Chapman and Cowling.[1] Figure 2 shows the dependence of $\alpha$ on $\omega$ based on Equations (19) and (21). From these equations, the values $\alpha=1$ and $\alpha=2.13986$ are obtained for hard-sphere and Maxwell molecules, respectively. It is emphasized that the values $\omega=1/2$ and $\alpha=1$ exactly reproduce the hard-sphere interaction (VSS and VHS are identical in this situation) but that the VSS values $\omega=1$ and $\alpha=2.13986$ (and the VHS values $\omega=1$ and $\alpha=1$) only approximate the IPL Maxwell interaction. Whenever a distinction is necessary, these Maxwell interactions are denoted as "VSS-Maxwell", "VHS-Maxwell", and "IPL-Maxwell". Although more computationally intense, the VSS interaction is emphasized here because it represents molecular diffusion more accurately than the VHS interaction and is thus appropriate for extending these investigations from single-species to multi-species gases.

CE theory provides the means to determine the relevant quantities for a particular IPL molecular interaction.[1] Table I contains numerical values for the infinite-to-first-approximation ratios of the viscosity, the thermal conductivity, and the self-diffusion coefficient ($\mu_\infty/\mu_1$, $K_\infty/K_1$, and $D_\infty/D_1$, respectively) and for the heat-flux and shear-stress Sonine-polynomial-coefficient ratios ($a_k/a_1$ and $b_k/b_1$) for the hard-sphere and Maxwell interactions.[17] Figure 2 also shows the dependence of $\mu_\infty/\mu_1$, $K_\infty/K_1$, and $D_\infty/D_1$ on $\omega$ for intermediate interactions. For the Maxwell interaction, the transport coefficient ratios are all unity, and the Sonine-polynomial-coefficient ratios are unity for $k=1$ and zero for $k \geq 2$. It is emphasized that these values are obtained only in the limit of small heat-flux and shear-stress Knudsen numbers.

## C. Effective transport coefficients

For Fourier flow and Couette flow, an effective thermal conductivity $K_{\text{eff}}$ and an effective viscosity $\mu_{\text{eff}}$ are defined as below:

$$q_x = -K_{\text{eff}} \frac{\partial T}{\partial x}, \qquad (22)$$

$$\tau_{xy} = \mu_{\text{eff}} \frac{\partial V}{\partial x}. \qquad (23)$$

When first-order CE theory applies (i.e., the heat-flux, shear-stress, and system Knudsen numbers are all small), Equations (8) and (9) indicate that $K_{\text{eff}} = K$ and $\mu_{\text{eff}} = \mu$. When first-order CE theory does not apply (i.e., at least one Knudsen number is finite), it is generally the case that $K_{\text{eff}} \neq K$ and $\mu_{\text{eff}} \neq \mu$. If the system Knudsen number is small but at least one of the other two Knudsen numbers is finite, the effective transport coefficient ratios $K_{\text{eff}}/K$ and $\mu_{\text{eff}}/\mu$, as well as the Sonine-polynomial-coefficient ratios $a_k/a_1$ and $b_k/b_1$ defined by Equations (13) and (14), are in general nonlinear functions of both $\text{Kn}_q$ and $\text{Kn}_\tau$.

### III. MOMENT-HIERARCHY METHOD

### A. Application to Maxwell molecules

Most of the known analytical solutions for the Boltzmann equation (BE) consider Maxwell molecules and apply the moment-hierarchy (MH) method directly or indirectly. Ikenberry and Truesdell[2] obtain an exact expression for the pressure tensor in uniform shear flow with Maxwell molecules. Asmolov et al.[3] and Makashev and Nosik[4] indirectly use the MH method to obtain expressions for higher-order moments for Fourier flow and Couette flow with Maxwell molecules. Santos and co-workers[5-7] use the MH method extensively to investigate Fourier flow and Couette flow with Maxwell molecules. Sabbane and Tij[8] present efficient computational algorithms that utilize symbol-manipulation software to calculate the collisional moments for Maxwell molecules that are required in the MH method.

The MH method is particularly useful for Maxwell molecules because the collision rate for the Maxwell interaction is independent of the relative speed of the molecules. This property allows the BE to be represented as an infinite hierarchy of moment equations.[5-7] The BE describes the temporal and spatial variation of the velocity distribution function $f$ and has the following form in the absence of body forces:

$$\partial f / \partial t + \mathbf{u} \cdot \nabla f = J[\mathbf{c} \,|\, f, f], \tag{24}$$

where $J[\mathbf{c} \,|\, f, f]$ is the collision operator. Moments of the BE relate moments of $f$ to moments of $J[\mathbf{c} \,|\, f, f]$, where their nondimensional forms are given below and $\tilde{f} = (c_m^3 / n) f$ :

$$M_{k_1 k_2 k_3} = \int \tilde{c}_x^{k_1} \tilde{c}_y^{k_2} \tilde{c}_z^{k_3} \tilde{f}[\tilde{\mathbf{c}}] d\tilde{\mathbf{c}} = \left\langle \tilde{c}_x^{k_1} \tilde{c}_y^{k_2} \tilde{c}_z^{k_3} \right\rangle, \tag{25}$$

$$J_{k_1 k_2 k_3} = \int \tilde{c}_x^{k_1} \tilde{c}_y^{k_2} \tilde{c}_z^{k_3} J[\tilde{\mathbf{c}} \,|\, \tilde{f}, \tilde{f}] d\tilde{\mathbf{c}}. \tag{26}$$

For the special situation of Maxwell molecules, the $J_{k_1 k_2 k_3}$ can be expressed as bilinear combinations of the $M_{k_1 k_2 k_3}$, where the coefficients in these combinations are linear combinations of the eigenvalues of the linearized collision operator.[5-7] These coefficients can be computed by the algorithm of Sabbane and Tij[8] and differ for the VSS-Maxwell, VHS-Maxwell, and IPL-Maxwell interactions because these interactions produce different angular scattering.

## B. Fourier flow and Couette flow

The above property enables an exact solution to the BE to be obtained recursively for Fourier flow and Couette flow with Maxwell molecules.[3-7] In this solution, the pressure $p = n k_B T$ is uniform in space, and the moments $M_{k_1 k_2 k_3}$ are polynomials of degree $k_1 + k_2 + k_3 - 2 \geq 0$ in the heat-flux Knudsen number $\mathrm{Kn}_q$:

$$M_{k_1 k_2 k_3}[\mathrm{Kn}_q, \mathrm{Kn}_\tau] = \sum_{j=0}^{k_1+k_2+k_3-2} \mu_{k_1 k_2 k_3}^{(j)}[\mathrm{Kn}_\tau] \mathrm{Kn}_q^j. \tag{27}$$

Here, the coefficients $\mu_{k_1 k_2 k_3}^{(j)}$ are nonlinear functions of the shear-stress Knudsen number $\mathrm{Kn}_\tau$ that can be represented by infinite expansions in powers of $\mathrm{Kn}_\tau$. Several of the lower-order moments have obvious values (number density, velocity, temperature, shear stress, heat flux):

$$M_{000} = 1, \tag{28}$$

$$M_{100} = M_{010} = M_{001} = 0, \tag{29}$$

$$M_{200} + M_{020} + M_{002} = 3/2, \tag{30}$$

$$M_{110} = -\mathrm{Kn}_\tau, \tag{31}$$

$$M_{300} + M_{120} + M_{102} = 2\mathrm{Kn}_q. \tag{32}$$

In the limit of small shear stress (i.e., $\mathrm{Kn}_\tau \to 0$), Equations (13), (14), (25), and (27) indicate that the Sonine-polynomial-coefficient ratios $a_k/a_1$ and $b_k/b_1$ for $k \geq 2$ are even polynomials of degree $2(k-1)$:

$$\frac{a_k}{a_1} = (-1)^{k-1} \sum_{j=1}^{k-1} A_{kj} \mathrm{Kn}_q^{2j}, \tag{33}$$

$$\frac{b_k}{b_1} = (-1)^{k-1} \sum_{j=1}^{k-1} B_{kj} \mathrm{Kn}_q^{2j}. \tag{34}$$

Table II contains values for the nonzero coefficients in the above equations for the IPL-Maxwell, VSS-Maxwell, and VHS-Maxwell interactions determined using the approach of Sabbane and Tij.[8] The differences between the values are caused by differences in the angular scattering. Note that $A_{51} = 0$. In general, $A_{kj} = 0$ if $j < (k-1)/3$.

## C. Transport coefficients

For Fourier flow and Couette flow under the assumption that a hydrodynamic description is valid (small system Knudsen number), the effective thermal conductivity and the effective viscosity defined by Equations (22) and (23) can potentially be functions of the heat-flux and shear-stress Knudsen numbers. For Fourier flow, Asmolov et al.[3] prove that the thermal conductivity for Maxwell molecules is independent of the heat flux. For Couette flow, Makashev and Nosik[4] prove that the viscosity for Maxwell molecules depends on the shear stress. Santos and co-workers[5-7] apply the MH method to determine the manner in which the thermal conductivity $K_{\mathrm{eff}}$ and the viscosity $\mu_{\mathrm{eff}}$ for Maxwell molecules (see Equations (22) and (23)) depend on the small but finite shear-stress Knudsen number, where $K$ and $\mu$ are the CE values:

$$K_{\mathrm{eff}}/K = F_K[\mathrm{Kn}_\tau] = 1 - c_K \mathrm{Kn}_\tau^2 + \mathrm{O}[\mathrm{Kn}_\tau^4], \tag{35}$$

$$\mu_{\mathrm{eff}}/\mu = F_\mu[\mathrm{Kn}_\tau] = 1 - c_\mu \mathrm{Kn}_\tau^2 + \mathrm{O}[\mathrm{Kn}_\tau^4]. \tag{36}$$

Table II shows the values of $c_K$ and $c_\mu$ for IPL-Maxwell, VSS-Maxwell, and VHS-Maxwell molecules. Since both coefficients are positive, Maxwell-molecule gases are shear-insulating and shear-thinning: the thermal conductivity and the viscosity decrease as the shear stress increases.

## IV. DSMC METHOD

## A. General description

The Direct Simulation Monte Carlo (DSMC) method of Bird[11] provides an additional method for investigating the behavior of gases under high heat flux and high shear stress.[13-17] DSMC uses computational molecules that move, reflect from walls, and collide with each other to simulate noncontinuum gas behavior. Each computational molecule typically represents a large number of real molecules. During a time step, molecules move at constant velocity ("ballistic" movement). Molecules that cross a solid boundary are reflected back into the computational domain. These reflections can be specular, diffuse at the wall temperature, diffuse without energy change, or a linear combination of these (in a probabilistic sense). More complicated reflection models are also available.[19] Between moves, pairs of molecules within each cell are randomly selected to collide at the appropriate rate.

The collisions of computational molecules mimic the collisions of real molecules statistically ("stochastic" collisions). Here, the variable-soft-sphere (VSS) interaction of Koura and Matsumoto[18] and the variable-hard-sphere (VHS) model of Bird[11] are used. As discussed above, these interactions exactly represent the hard-sphere interaction and approximate other inverse-power-law (IPL) interactions, including the Maxwell interaction, when suitable values are selected for the viscosity temperature exponent $\omega$ and the angular-scattering exponent $\alpha$. Although slightly more expensive computationally, the VSS interaction is emphasized here because it represents molecular diffusion more accurately than the VHS interaction.

The computational mesh in DSMC serves two functions. First, the computational mesh enables identification of pairs of molecules as possible collision partners. Second, it provides a means for accumulating statistical information about the flow (e.g., the number density, the velocity, the temperature, the shear stress, the heat flux, and other moments of the velocity distribution function). Statistical information is sampled both before and after collisions are performed. If the flow is statistically steady (stationary), long-time averages are used to reduce statistical uncertainty (the ergodic hypothesis).

## B. Simulation specifics

The flow domain shown in Figure 1 is considered. Table III shows the physical parameters used in the simulations. As in previous studies,[15-17] the gas has the molecular mass and the reference viscosity of argon. However, the $\omega$ and $\alpha$ values for the VSS interaction are used as discussed above to represent hard-sphere, Maxwell, and intermediate IPL interactions. Initially, the gas is motionless and at the reference pressure and temperature: $p_{init} = p_{ref} = 266.644 \text{ Pa}$ (2 torr) and $T_{init} = T_{ref} = 273.15 \text{ K}$, respectively. The most probable molecular thermal speed at these conditions is $c_m = 337.3 \text{ m/s}$. The domain has a length $L = 1 \text{ mm}$ and is bounded by two parallel solid walls that reflect all molecules diffusely at the wall temperature (unity accommodation). These walls have temperatures $T_1 = T_{ref} - \Delta T/2$ and $T_2 = T_{ref} + \Delta T/2$, respectively, and tangential velocities $V_1 = -\Delta V/2$ and $V_2 = \Delta V/2$, respectively. Temperature differences up to $\Delta T = 400 \text{ K}$ and velocity differences up to $\Delta V = 800 \text{ m/s}$ are considered. With the mean free path defined as

$$\lambda = \frac{\sqrt{\pi}\mu}{mnc_m} = \frac{\sqrt{\pi}\mu c_m}{2p}, \tag{37}$$

the system Knudsen number at the initial conditions has the value $\lambda/L = 0.0237$, so the walls are about 42 mean free paths apart. Since the Knudsen layers produced by the walls are generally about 4-10 mean free paths thick, the normal solution occupies a large fraction of the domain.

A slightly modified version of the code DSMC1 is used to perform these simulations. The principal modification from the published version of the code[11] involves performing both pre-collision and post-collision sampling, which improves the accuracy of non-conserved moments.[15-17] Table III shows the numerical parameters for the simulations. The domain is divided into 400 uniform cells of width $\Delta x = 2.5 \ \mu$m (about 1/10 of the mean free path $\lambda$) and is populated with $N_c = 120$ computational molecules per cell selected from an equilibrium distribution at the initial pressure and temperature. A fixed time step of $\Delta t = 7$ ns (about 1/10 of the collision time $\lambda/c_m$) is used. Transients are allowed to decay for the first 0.3 million time steps (2.1 ms), and approximately 1 billion samples per cell are subsequently accumulated. Based on extensive simulations, discretization and stochastic errors in the thermal conductivity and the viscosity are expected to be below 0.2%.[16] Each simulation requires 60-120 hours on 16 nodes of an IBM Linux cluster with dual 1.2-GHz P3 processors.

Quantities of interest include the thermal conductivity, the viscosity, and the Sonine-polynomial coefficients. The effective thermal conductivity $K_{\text{eff}}$ and the effective viscosity $\mu_{\text{eff}}$ are determined within each mesh cell from Equations (17), (18), (22), and (23):

$$\frac{K}{K_{\text{eff}}} = -\frac{K}{q_x}\frac{\partial T}{\partial x} = -\frac{K_{\text{ref}}}{T_{\text{ref}}^{\omega}}\frac{T^{\omega}}{q_x}\frac{\partial T}{\partial x}, \tag{38}$$

$$\frac{\mu}{\mu_{\text{eff}}} = \frac{\mu}{\tau_{xy}}\frac{\partial V}{\partial x} = \frac{\mu_{\text{ref}}}{T_{\text{ref}}^{\omega}}\frac{T^{\omega}}{\tau_{xy}}\frac{\partial V}{\partial x}, \tag{39}$$

where the reference quantities are known, the quantities $T$, $V$, $q_x$, and $\tau_{xy}$ are determined within each cell, and the spatial derivatives $\partial T/\partial x$ and $\partial V/\partial x$ are approximated using central differences with values from adjacent cells. The moments needed to determine the $a_k/a_1$ and the $b_k/b_1$ from Equations (13) and (14) are accumulated directly within each cell during a simulation. The above quantities are examined as functions of position (profiles) or as functions of the heat-flux or shear-stress Knudsen number in the central region of the domain, within which the normal solution is obtained. When relevant, average values are also obtained for properties in this region.

## V. RESULTS

### A. Weakly nonequilibrium conditions

Before considering highly nonequilibrium conditions, weakly nonequilibrium conditions are first examined. The reasons for this are twofold. First, the accuracy of the DSMC method is demonstrated by simulating conditions for which first-order CE theory applies. This level of agreement provides confidence in the DSMC method when theoretical results are not available. Second, the region of the domain occupied by the normal solution and the complementary region occupied by the Knudsen layers are determined. The fact that the normal solution is obtained in a significant portion of the domain for these simulations provides confidence that the normal solution is obtained over a similar region at highly nonequilibrium conditions.

Figure 3 presents the temperature and velocity profiles for a simulation with Maxwell molecules for $\Delta T = 70$ K and $\Delta V = 100$ m/s. In this simulation, the left wall is colder ($T_1 = 238.15$ K) and moving downward ($V_1 = -50$ m/s), whereas the right wall is hotter ($T_2 = 308.15$ K) and moving upward ($V_2 = 50$ m/s). The heat-flux and shear-stress Knudsen numbers corresponding to these conditions are $Kn_q \approx 0.006$ and $Kn_\tau \approx 0.003$, respectively, so CE theory is expected to apply in the central region of the domain. Due to the large number of samples obtained in the simulation, the statistical variations in these profiles cannot be visually discerned. The discontinuities between the gas and wall values of the temperature and the velocity are small (about 3 K and 2 m/s, respectively) because of the small system Knudsen number (0.0237). Outside the Knudsen layers (10-25% of the domain adjacent to each wall), the temperature and velocity profiles are nearly linear with a slight downward concavity. This shape results from two effects. First, the thermal conductivity and the viscosity are proportional to the temperature for Maxwell molecules (they are increasing functions of temperature for IPL, VSS, and VHS molecules), so the transport coefficients are larger on the right side of the domain than on the left side. Since the heat flux and the shear stress are uniform across the domain, the temperature and velocity gradients are smaller on the right side than on the left side. Second, viscous dissipation generates heat throughout the domain. As discussed earlier, this phenomenon acts to create a maximum in temperature in the center of the domain even when both walls are at the same temperature. The temperature increase is small compared to the wall temperatures for the above velocity difference (about 2 K for $\Delta V = 100$ m/s ) but can be much larger for large velocity differences (e.g., over 100 K for $\Delta V = 800$ m/s ).

Figure 4 shows the normalized effective thermal-conductivity and viscosity profiles for these conditions. The effective values $K_{eff}$ and $\mu_{eff}$ are determined using Equations (22) and (23) and are normalized using the CE values, yielding Equations (38) and (39). It is noted that the CE thermal conductivity $K$ and viscosity $\mu$ are not constant within the domain but rather increase from left to right because they are increasing functions of temperature and the temperature increases monotonically (in this case) from left to right. A value of unity indicates that the CE value is obtained, which occurs in the central region of the domain for both transport coefficients. The stochastic noise evident in these plots enters primarily through the numerical derivatives used in Equations (38) and (39). The Knudsen layers are clearly evident and are restricted to about 10-25% of the domain adjacent to each wall. The Knudsen layer near the hot wall is slightly thicker than the Knudsen layer near the cold wall because the mean free path at constant pressure is an increasing function of temperature (see Equations (37) and (18)).

Figure 5 shows the profiles of the Sonine-polynomial-coefficient ratios $a_k/a_1$ and $b_k/b_1$ for these conditions. The $a_k/a_1$ are shown for $k = 2, 3, 4,$ and $5$, and the $b_k/b_1$ are shown for $k = 2, 3,$ and $4$. The solid curves are the DSMC values from Equations (13) and (14), and the dashed lines indicate the CE values of 0 that are obtained for Maxwell molecules when $k \geq 2$. As observed regarding the thermal conductivity and the viscosity, the $a_k/a_1$ and the $b_k/b_1$ achieve their CE values in the central region of the domain and depart from the CE values only in the Knudsen layers, which again are seen to occupy about 10-25% of the domain adjacent to the walls. As above, the normal solution is seen to occupy the central region of the domain.

Figure 6 shows the corresponding profiles obtained using hard-sphere molecules instead of Maxwell molecules but with all other conditions unchanged. The profiles of temperature, velocity, thermal conductivity, and viscosity for hard-sphere molecules are almost identical to the corresponding profiles for Maxwell molecules in Figures 3-4 (with slight differences arising from the different temperature dependence of the transport coefficients), so these profiles are not shown. As for the simulation using Maxwell molecules, the heat-flux and shear-stress Knudsen numbers at these conditions are $\mathrm{Kn}_q \approx 0.006$ and $\mathrm{Kn}_\tau \approx 0.003$, respectively, so CE theory is expected to apply in the central region of the domain. The CE values are also shown as in the previous figure. Unlike Maxwell molecules, hard-sphere molecules have nonzero CE values for the $a_k/a_1$ and the $b_k/b_1$ when $k \geq 2$ (see Table I). As with Maxwell molecules, CE behavior is obtained in the central region of the domain, and departures are observed only in the Knudsen layers adjacent to the walls.

Figure 7 shows the normalized effective thermal conductivity and viscosity as functions of the viscous temperature exponent for IPL interactions from hard-sphere through Maxwell. These values are obtained by averaging profiles like those in Figure 4 over the central region of the domain, with error bars indicating the 95% confidence intervals corresponding to the stochastic uncertainty. Six $(\omega, \alpha)$ combinations are examined (see Figure 2 and Equations (19) and (21)): $(0.5, 1.0)$, $(0.6, 1.20904)$, $(0.7, 1.42248)$, $(0.8, 1.64556)$, $(0.9, 1.88313)$, and $(1.0, 2.13986)$, where the first combination is the hard-sphere interaction and the last combination is the VSS-Maxwell interaction. In all cases except for the Maxwell interaction, the values for the thermal conductivity and the viscosity agree with the CE values to within the error bars, which are about $\pm 0.002$. As indicated earlier, the discretization errors for these simulations are also about $\pm 0.002$ and tend to be positive for the hard-sphere interaction but negative for the Maxwell interaction, as shown below. The high accuracy of these values is further emphasized by recognizing that the differences between the infinite-approximation and first-approximation CE values of the thermal conductivity and the viscosity for hard-sphere molecules are about 0.025 and 0.016, respectively (see Table I), which are roughly 10 times as large as the differences exhibited in this figure.

Figure 8 shows the Sonine-polynomial-coefficient ratios as functions of the viscosity temperature exponent for the conditions in the previous figure. As in the previous figure, these values are obtained by averaging over the central region. However, error bars are not shown because the stochastic errors are smaller than the symbol size. As in the previous figures, the dashed curves are the CE values. Since the heat-flux and shear-stress Knudsen numbers ($\mathrm{Kn}_q \approx 0.006$ and $\mathrm{Kn}_\tau \approx 0.003$, respectively) are small in these simulations, the good agreement between the simulation values in the central region of the domain and the CE values is not surprising. Close examination of this figure indicates that the coefficient ratios actually depart slightly but systematically from the CE values (the most noticeable example is $a_3/a_1$). This is because the heat-flux Knudsen number is not infinitesimally small. Simulations with smaller temperature differences corroborate this observation.[15]

To summarize, when the heat-flux and shear-stress Knudsen numbers are small, the DSMC simulations exhibit CE behavior in the central region of the domain. More specifically, the CE values for the thermal conductivity and the viscosity (both temperature-dependent) and for the Sonine-polynomial-coefficient ratios are obtained to high accuracy in this region, indicating that DSMC correctly reproduces the CE velocity distribution. The Knudsen layers are seen to occupy about 10-25% of the domain adjacent to each wall. Thus, the normal solution of the BE is obtained in the central region of the domain. Since the mean free path is an increasing function of temperature at constant pressure, the Knudsen layer adjacent to the right wall is thicker than the Knudsen layer adjacent to the left wall. This leftward shift of the region in which the normal solution is obtained becomes more pronounced at larger temperature differences. These observations hold for all interactions from hard-sphere through Maxwell. Additional DSMC results using the VHS interaction (not shown) also exhibit CE behavior in the central region and are nearly identical to those presented above using the VSS interaction, which is as expected for single-species gases because the effect of molecular diffusion is not significant. Finally, close examination of the DSMC results indicates slight departures from CE theory that are attributed to the fact that the heat-flux Knudsen number is not infinitesimally small.

## B. Highly nonequilibrium conditions

With highly nonequilibrium conditions, the CE theory discussed in earlier sections is no longer applicable. For Maxwell molecules, first-order CE theory is superseded by the MH theory, which provides expressions for the effective thermal conductivity, the effective viscosity, and the Sonine-polynomial-coefficient ratios in terms of the heat-flux and shear-stress Knudsen numbers. For hard-sphere and other IPL molecules, no corresponding theory is available. Thus, for highly nonequilibrium conditions, the DSMC method is verified by comparison with MH results for Maxwell molecules and provides new information for hard-sphere molecules.

Figure 9 shows the profiles of the Sonine-polynomial-coefficient ratios for Maxwell molecules at the same conditions as in Figure 5 except that the temperature difference $\Delta T$ is increased from 70 K to 200 K. Under these conditions, the heat-flux Knudsen number $\text{Kn}_q$ is correspondingly increased from about 0.006 to about 0.017. The approximate nature of the preceding statement reflects the fact that the heat-flux Knudsen number is not constant throughout the domain but increases from hot to cold at constant pressure (i.e., from right to left in the simulations). This is because the heat-flux Knudsen number can be expressed as $\text{Kn}_q = |q_x|/(2pc_m)$, the heat flux is uniform in the domain, the pressure is approximately uniform in the domain (exactly so for the Maxwell-molecule normal solution[3-7]), and the temperature and the most probable molecular speed increase from left to right. This variation with position is quite small for the previous simulation with $\Delta T = 70$ K : $c_m$ and $\text{Kn}_q$ lie within 7% of their average values everywhere in the domain, which is the reason for ignoring the spatial variation of $\text{Kn}_q$ when determining averages in the previous section. As in Figure 5, the solid curves are the DSMC results, and the dashed lines are the CE results. At these finite values of the heat-flux Knudsen number, the Knudsen layers and the central region are still evident although the central region is shifted leftward to account for the temperature dependence of the mean free path at constant pressure. However, the DSMC results differ significantly from the CE results in the central region of the domain. More specifically, the Sonine-polynomial-coefficient ratios differ increasingly from the corresponding CE values from right to left just as the heat-flux Knudsen number increases from right to left. Within the central region, the variation of the $a_k/a_1$ and the $b_k/b_1$ with $Kn_q$ represents the normal solution. Thus, a single DSMC simulation provides the normal solution for the $Kn_q$ values in the central region. The same approach is used subsequently to determine the variation of the normalized effective thermal conductivity and viscosity with the heat-flux Knudsen number. However, as shown below, these quantities remain so close to unity that their values, along with the heat-flux Knudsen number, are averaged over the central region to reduce stochastic errors.

Figure 10 shows the Sonine-polynomial-coefficient ratios for Maxwell molecules as functions of the heat-flux Knudsen number as determined in the manner above. The symbols indicate the values determined from DSMC simulations. Each cluster of points along a curve corresponds to values obtained from the central region of a single DSMC simulation as discussed above. More specifically, the four clusters correspond to four DSMC simulations with temperature differences of $\Delta T = 70$, 200, 300, and 400 K and a velocity difference of $\Delta V = 100 \text{ m/s}$. In all cases, the shear-stress Knudsen number is below 0.005, which is small. The solid and long-dashed curves are the corresponding MH results for VSS-Maxwell and IPL-Maxwell interactions, respectively, in the zero-shear-stress limit ($Kn_\tau \to 0$), as given by Equations (33) and (34) with the parameter values in Table II, and the dashed lines indicate the CE values of 0. The DSMC values are seen to agree closely with the MH VSS-Maxwell values except for $a_4/a_1$ and $a_5/a_1$ at $\Delta T = 400 \text{ K}$ (the largest temperature difference), whereas the DSMC VSS-Maxwell values differ increasingly from the MH IPL-Maxwell values as $k$ increases. The former slight difference between the DSMC and VSS-Maxwell values appears to have two causes. First, discretization errors appear to account for about half of this difference based on additional DSMC simulations in which $\Delta x$ and $\Delta t$ are halved while $N_c$ is doubled (the stochastic errors increase substantially, so a more detailed quantification is not possible). Second, the finite shear-stress Knudsen number in the DSMC simulations also accounts for about half of this difference based on additional simulations at a shear-stress Knudsen number that is approximately twice as large (i.e., a velocity difference of $\Delta V = 200$ m/s ).

Figure 11 shows the Sonine-polynomial-coefficient ratios for hard-sphere molecules as functions of the heat-flux Knudsen number at the same conditions as in the previous figure. The long-dashed curves are low-order polynomial fits to the DSMC values (there are no theoretical results available for comparison), and the dashed lines are the CE values. The dependence on the heat-flux Knudsen number is similar for hard-sphere and Maxwell molecules. More specifically, coefficient ratios with even values of $k$ decrease with increasing $\mathrm{Kn}_q$, whereas coefficient ratios with odd values of $k$ increase with increasing $\mathrm{Kn}_q$. However, the rate of change is more gradual for hard-sphere molecules than it is for Maxwell molecules. Unlike the DSMC results for VSS-Maxwell molecules, the DSMC results for hard-sphere molecules are almost independent of the discretization parameters $\Delta x$, $\Delta t$, and $N_c$ and of the shear-stress Knudsen number $\mathrm{Kn}_\tau$. It is conjectured that this difference may be related to the fact that the molecular collision rate is an increasing function of the relative molecular speed for hard-sphere molecules but is independent of the relative molecular speed for Maxwell molecules, which allows high-speed Maxwell molecules to travel farther than their hard-sphere counterparts before colliding.

Figure 12 shows the normalized effective thermal conductivity and viscosity for Maxwell molecules as functions of the heat-flux Knudsen number for several values of the shear-stress Knudsen number. The four solid symbols along each of the three curves (drawn to guide the eye) in the thermal-conductivity plot are values from four DSMC simulations identical to those presented in Figure 10 except with velocity differences of $\Delta V = 0,\ 100,\ $ and $200\ \text{m/s}$. Since the shear stress must be nonzero in order to determine the effective viscosity, only values from the latter two simulations are present in the viscosity plot. In distinction to Figure 10, the DSMC values in Figure 12 are averaged over the central region of the domain, with error bars corresponding to the 95% confidence intervals. Averaging of these ratios is acceptable because their spatial variation in the central region of the domain is quite small (the profiles of these quantities are similar to those in Figure 4). The dashed lines indicate the CE and MH results for Maxwell molecules. As the shear-stress Knudsen number approaches zero, the DSMC values increase upward toward the MH values. The DSMC values exhibit a slight but consistent increase with increasing heat-flux Knudsen number. However, the net increase over the entire range of heat-flux Knudsen number is comparable to the stochastic and discretization errors for these simulations, which are each about $\pm 0.002$. Thus, to within numerical uncertainty, the Maxwell-molecule transport coefficients are independent of heat flux, in agreement with theory.[3-7]

Figure 13 shows the normalized effective thermal conductivity and viscosity for hard-sphere molecules as functions of the heat-flux Knudsen number at the same conditions as in the previous figure. In this figure, the dashed lines represent the CE values, which are relevant only in the limit of zero heat-flux Knudsen number. No theoretical results for finite heat-flux Knudsen number are available for comparison. As observed for Maxwell molecules, the hard-sphere values increase upward as the shear-stress Knudsen number is decreased, and this variation is nearly linear. However, the CE values are not obtained except for small values of the heat-flux Knudsen number. Instead, the normalized effective thermal conductivity and viscosity decrease approximately quadratically with increasing heat-flux Knudsen number. The differences observed at the largest heat-flux Knudsen number are slightly larger than the combined effect of discretization and stochastic errors (about $\pm 0.002$ each) and thus appear to be real. As noted above, the DSMC results for hard-sphere molecules are almost independent of the discretization parameters $\Delta x$, $\Delta t$, and $N_c$, which provides further evidence supporting this assertion.

Figure 14 shows the normalized effective thermal conductivity and viscosity for Maxwell molecules as functions of the shear-stress Knudsen number. The symbols represent values from DSMC simulations with tangential velocity differences of $\Delta V = 0$-800 m/s in increments of 100 m/s and a temperature difference of $\Delta T = 0$ K. Even though the walls are at the same temperature, the effective thermal conductivity can be determined because viscous dissipation produces spatially nonuniform temperature and heat-flux profiles. (A small region centered around the midpoint of the domain is excluded because the temperature gradient and the heat flux are too small to determine the effective thermal conductivity with acceptable stochastic uncertainty.) As in previous figures, the error bars represent the 95% confidence intervals. In the thermal-conductivity plot, the error bars are seen to increase as the shear-stress Knudsen number is decreased. This phenomenon results from the fact that the temperature gradients and the heat flux produced by viscous heating are approximately proportional to the square of the shear-stress Knudsen number, whereas the stochastic noise does not depend strongly on the shear-stress Knudsen number. In the viscosity plot, the error bars do not depend strongly on the shear-stress Knudsen number because the shear stress is essentially proportional to the velocity gradient and the ratio of the shear stress to the velocity gradient determines the effective viscosity. Also shown in this figure are three theoretical curves. The dashed lines are the CE results, which are appropriate only for vanishing shear-stress and heat-flux Knudsen numbers, the solid curves are the MH results for Maxwell molecules (Equations (35) and (36)), and the long-dashed curves are the MH results offset by a small arbitrary amount. The offset MH results are in excellent agreement with the DSMC results. The small negative offsets ($-0.002$ for both transport coefficients) are comparable to and presumably represent the discretization errors discussed above ($\pm 0.002$). Throughout the entire domain in all of these DSMC simulations, the heat-flux Knudsen number is below 0.017; in fact, it is well below this value for all but the largest velocity differences. Based on Figure 12, the heat-flux Knudsen number does not affect the normalized effective thermal conductivity and viscosity significantly in these simulations.

Figure 15 shows the normalized effective thermal conductivity and viscosity for hard-sphere molecules as functions of the shear-stress Knudsen number at the same conditions as in the previous figure. The symbols represent the DSMC values, the dashed lines represent the CE values, which are appropriate only for vanishing shear stress, and the long-dashed curves represent curve fits of the form $c_0 + c_2 \mathrm{Kn}_\tau^2$ through the DSMC values. There are no exact theoretical results available for comparison. The curve fits are seen to represent all of the DSMC values accurately. The intercepts of these curves are positive and lie between 0.001 and 0.002. These values are comparable to and presumably represent the discretization errors ($\pm 0.002$). The rightward extrapolations of these curves are shown simply to guide the eye and should be quantitatively assessed prior to use in another context. When compared to Maxwell molecules (see Figure 14), hard-sphere molecules exhibit a weaker dependence of the normalized effective thermal conductivity and viscosity on the shear-stress Knudsen number than Maxwell molecules do. Nevertheless, the departures of these quantities from unity at large shear-stress Knudsen numbers are significantly larger than the discretization error, the stochastic error, and their combined effect. Coincidentally, these departures are similar to the differences between the infinite-approximation and first-approximation hard-sphere CE values for the thermal conductivity and the viscosity, namely 0.025 and 0.016 as in Table I.

To summarize, DSMC simulations are presented to determine the behavior of the normal solution when either the heat-flux Knudsen number or the shear-stress Knudsen number is finite. The DSMC and MH results for the VSS-Maxwell interaction are in very good agreement (and are clearly different from the MH IPL-Maxwell results), which provides strong evidence that DSMC produces the correct velocity distribution. While not shown, the DSMC and MH VHS-Maxwell results are nearly identical to the corresponding VSS-Maxwell results, as expected in situations where the effect of molecular diffusion is not significant. No exact theoretical hard-sphere results are available, but the DSMC hard-sphere and Maxwell results exhibit similar behavior except in one particular. Although the normalized effective thermal conductivity and viscosity are decreasing functions of the shear-stress Knudsen number for both Maxwell and hard-sphere molecules, these quantities are essentially independent of the heat-flux Knudsen number for Maxwell molecules whereas they are both decreasing functions of the heat-flux Knudsen number for hard-sphere molecules. It is noted that the decreases in these quantities observed in the DSMC simulations presented above are always less than 3%. Thus, even under highly nonequilibrium conditions, the CE values for these quantities are probably accurate enough for many engineering calculations. These results are in accord with those of Santos and co-workers, who report similar agreement between DSMC simulations and theoretical results for BGK-like collision terms for much larger heat-flux and shear-stress values.[13-14]

## VI. CONCLUSIONS

The state of a single-species monatomic gas experiencing a large heat flux or a large shear stress is investigated using the moment-hierarchy (MH) method for the Maxwell molecular interaction and using the Direct Simulation Monte Carlo (DSMC) method of Bird for Maxwell, hard-sphere, and intermediate molecular interactions. Normal solutions of the Boltzmann equation are found for Fourier flow (uniform heat flux) and Couette flow (uniform shear stress) for finite heat-flux and shear-stress Knudsen numbers. The thermal conductivity, the viscosity, and the Sonine-polynomial coefficients from the MH and DSMC methods agree with Chapman-Enskog (CE) theory at small Knudsen numbers. Additionally, these quantities are in agreement at finite Knudsen numbers for VSS-Maxwell and VHS-Maxwell molecules, which yield nearly identical results when molecular diffusion is not significant. The MH and DSMC methods both indicate that the effective thermal conductivity and the effective viscosity for Maxwell molecules are almost independent of the heat-flux Knudsen number but decrease slightly as the shear-stress Knudsen number is increased. Additional DSMC simulations indicate that these transport properties for hard-sphere molecules decrease slightly as the shear-stress Knudsen number or the heat-flux Knudsen number is increased. In all cases examined, these decreases are less than 3%, which indicates that the CE values for the thermal conductivity and the viscosity can be used under highly nonequilibrium conditions with small errors so long as the system Knudsen number is small (the hydrodynamic limit). These results provide strong evidence that the DSMC method can be used to determine the state of a gas under highly nonequilibrium conditions. Future work will involve using DSMC to investigate the behavior of multi-species gases for situations in which molecular diffusion is important (e.g., thermal diffusion).

**ACKNOWLEDGMENTS**

Sandia is a multiprogram laboratory operated by Sandia Corporation, a Lockheed Martin Company, for the United States Department of Energy's National Nuclear Security Administration under contract DE-AC04-94AL85000. The research of A. S. has been supported by the Ministerio de Educación y Ciencia (Spain) through Grant No. FIS2004-01399 (partially financed by FEDER funds).

TABLE I. Chapman-Enskog (CE) results.

| Symbol | Hard-Sphere | Maxwell |
|--------|-------------|---------|
| $\omega$ | 1/2 | 1 |
| $\alpha$ (VSS) | 1 | 2.13986 |
| $\alpha$ (VHS) | 1 | 1 |
| $\mu_\infty/\mu_1$ | 1.016034 | 1 |
| $K_\infty/K_1$ | 1.025218 | 1 |
| $D_\infty/D_1$ | 1.018954 | 1 |
| $a_1/a_1$ | 1 | 1 |
| $a_2/a_1$ | 0.0954284 | 0 |
| $a_3/a_1$ | 0.0217503 | 0 |
| $a_4/a_1$ | 0.0068579 | 0 |
| $a_5/a_1$ | 0.0025926 | 0 |
| $b_1/b_1$ | 1 | 1 |
| $b_2/b_1$ | 0.0617421 | 0 |
| $b_3/b_1$ | 0.0103303 | 0 |
| $b_4/b_1$ | 0.0025919 | 0 |
| $b_5/b_1$ | 0.0008207 | 0 |

TABLE II. Moment-hierarchy (MH) results.

| Symbol | IPL-Maxwell | VSS-Maxwell | VHS-Maxwell |
|--------|-------------|-------------|-------------|
| $A_{21}$ | 21.1786 | 21.3155 | 21.3190 |
| $A_{31}$ | 30.8947 | 32.4522 | 32.4917 |
| $A_{32}$ | 1455.17 | 1584.61 | 1588.02 |
| $A_{41}$ | 11.1479 | 12.0749 | 12.0990 |
| $A_{42}$ | 4539.44 | 5272.08 | 5291.97 |
| $A_{43}$ | 222458. | 288112. | 290023. |
| $A_{52}$ | 5514.96 | 6805.30 | 6841.24 |
| $A_{53}$ | $1.10275 \times 10^6$ | $1.55384 \times 10^6$ | $1.56741 \times 10^6$ |
| $A_{54}$ | $6.29132 \times 10^7$ | $1.06469 \times 10^8$ | $1.07906 \times 10^8$ |
| $B_{21}$ | 16.3894 | 16.5355 | 16.5392 |
| $B_{31}$ | 22.4570 | 24.1045 | 24.1472 |
| $B_{32}$ | 908.683 | 1007.93 | 1010.59 |
| $B_{41}$ | 6.61697 | 7.33624 | 7.35514 |
| $B_{42}$ | 2685.43 | 3210.81 | 3225.37 |
| $B_{43}$ | 115761. | 154863. | 156027. |
| $c_K$ | 29.0383 | 29.0418 | 29.0419 |
| $c_\mu$ | 596/45 | 596/45 | 596/45 |

TABLE III. DSMC simulation parameters.

| Quantity | Symbol | Value |
|---|---|---|
| Boltzmann constant | $k_B$ | $1.380658 \times 10^{-23}$ J/K |
| Molecular mass | $m$ | $66.3 \times 10^{-27}$ kg |
| Reference viscosity | $\mu_{\text{ref}}$ | $2.117 \times 10^{-5}$ Pa·s |
| Reference temperature | $T_{\text{ref}}$ | 273.15 K |
| Reference pressure | $p_{\text{ref}}$ | 266.644 Pa |
| Initial temperature | $T_{\text{init}}$ | $T_{\text{ref}}$ |
| Initial pressure | $p_{\text{init}}$ | $p_{\text{ref}}$ |
| Left wall temperature | $T_1$ | $T_{\text{ref}} - \Delta T / 2$ |
| Right wall temperature | $T_2$ | $T_{\text{ref}} + \Delta T / 2$ |
| Temperature difference | $\Delta T$ | Up to 400 K |
| Left wall velocity | $V_1$ | $-\Delta V / 2$ |
| Right wall velocity | $V_2$ | $\Delta V / 2$ |
| Velocity difference | $\Delta V$ | Up to 800 m/s |
| Domain length | $L$ | 1 mm |
| Cell size | $\Delta x$ | 2.5 μm |
| Time step | $\Delta t$ | 7 ns |
| Molecules per cell | $N_c$ | 120 |

## FIGURE CAPTIONS

FIG. 1.     Fourier flow and Couette flow.

FIG. 2.     Dependence of CE transport-coefficient ratios and VSS $\alpha$ on $\omega$.

FIG. 3.     Maxwell temperature and velocity profiles at small $\mathrm{Kn}_q$.

FIG. 4.     Maxwell thermal-conductivity and viscosity profiles at small $\mathrm{Kn}_q$.

FIG. 5.     Maxwell Sonine-polynomial-coefficient profiles at small $\mathrm{Kn}_q$.

FIG. 6.     Hard-sphere Sonine-polynomial-coefficient profiles at small $\mathrm{Kn}_q$.

FIG. 7.     Dependence of thermal conductivity and viscosity on $\omega$ at small $\mathrm{Kn}_q$.

FIG. 8.     Dependence of Sonine-polynomial coefficients on $\omega$ at small $\mathrm{Kn}_q$.

FIG. 9.     Maxwell Sonine-polynomial-coefficient profiles at finite $\mathrm{Kn}_q$.

FIG. 10.    Dependence of Maxwell Sonine-polynomial coefficients on $\mathrm{Kn}_q$.

FIG. 11.    Dependence of hard-sphere Sonine-polynomial coefficients on $\mathrm{Kn}_q$.

FIG. 12.    Dependence of Maxwell thermal conductivity and viscosity on $\mathrm{Kn}_q$.

FIG. 13.    Dependence of hard-sphere thermal conductivity and viscosity on $\mathrm{Kn}_q$.

FIG. 14.    Dependence of Maxwell thermal conductivity and viscosity on $\mathrm{Kn}_\tau$.

FIG. 15.    Dependence of hard-sphere thermal conductivity and viscosity on $\mathrm{Kn}_\tau$.

FIG. 1.

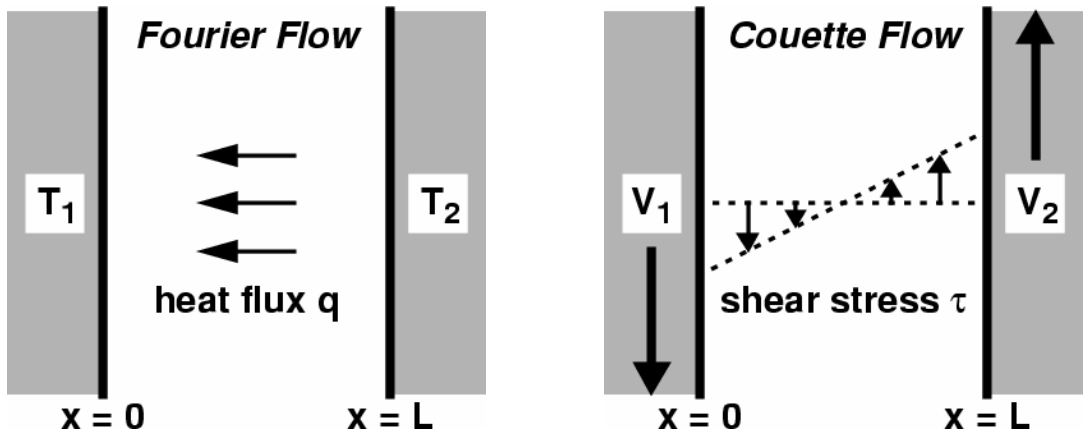

FIG. 2.

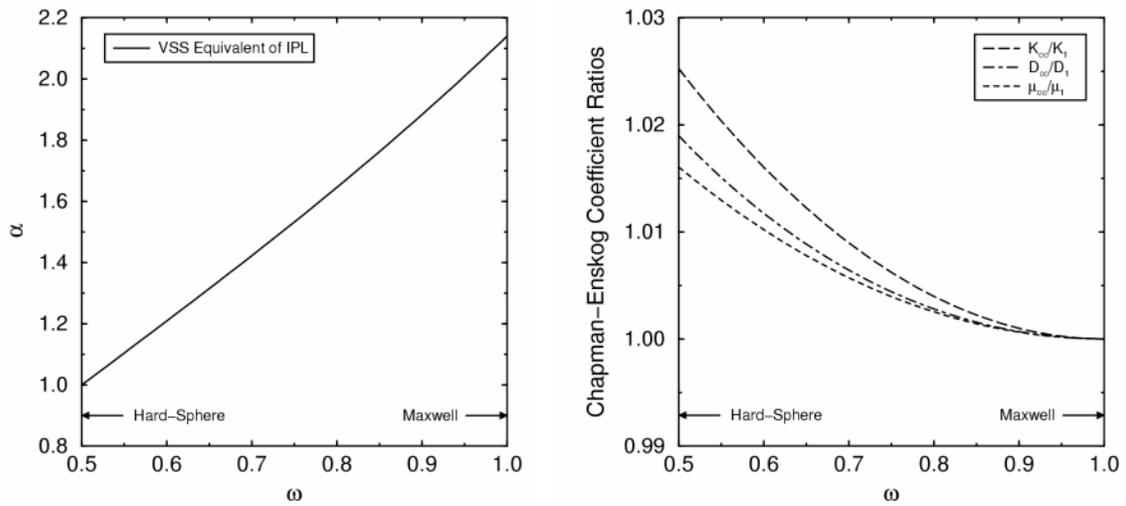

FIG. 3.

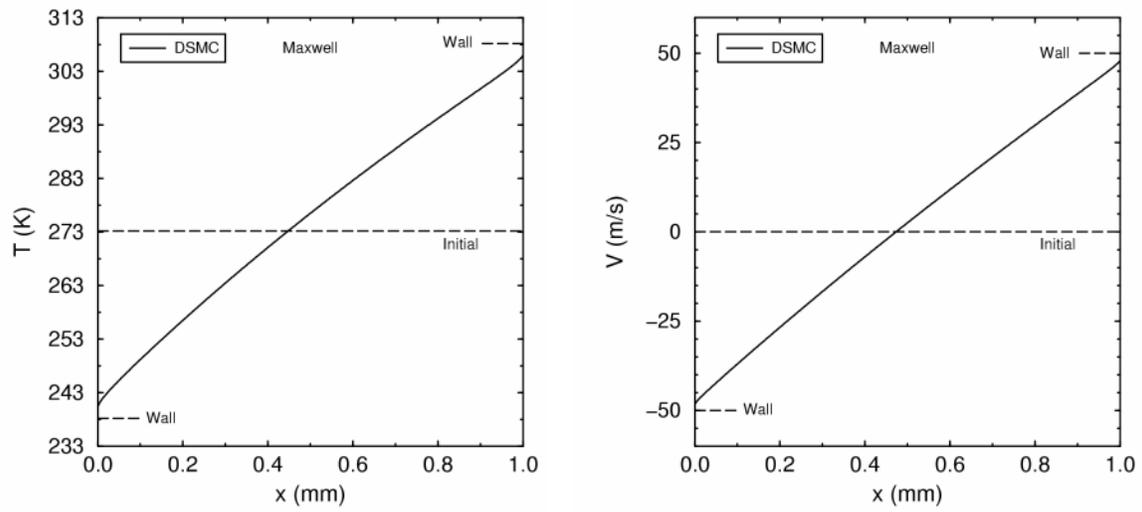

FIG. 4.

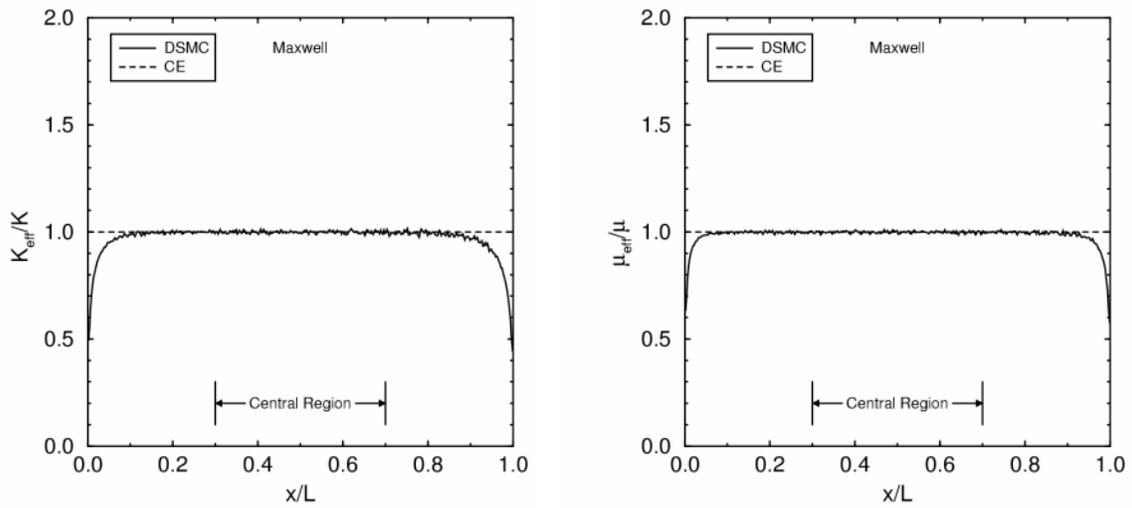

FIG. 5.

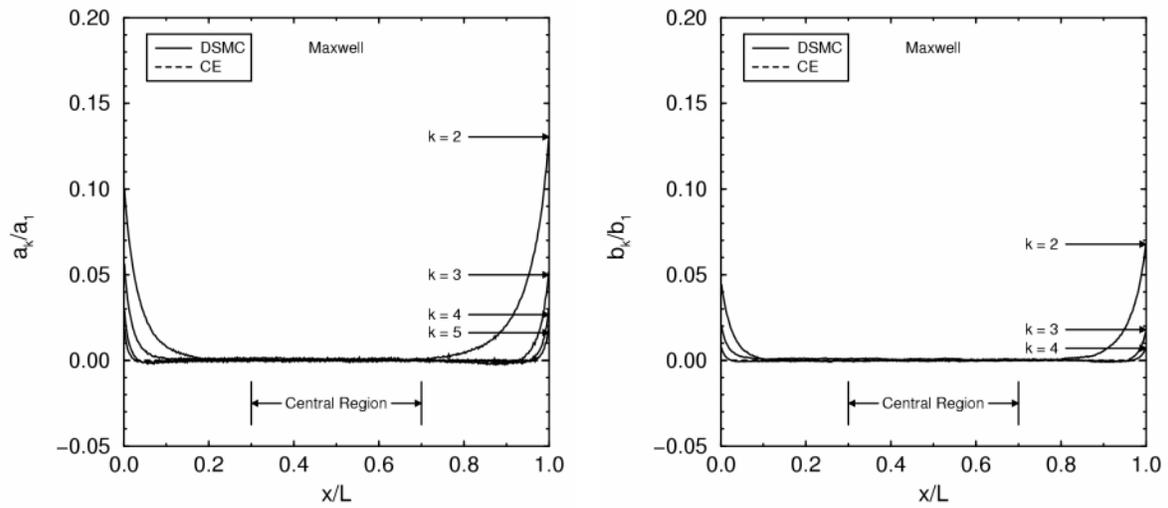

FIG. 6.

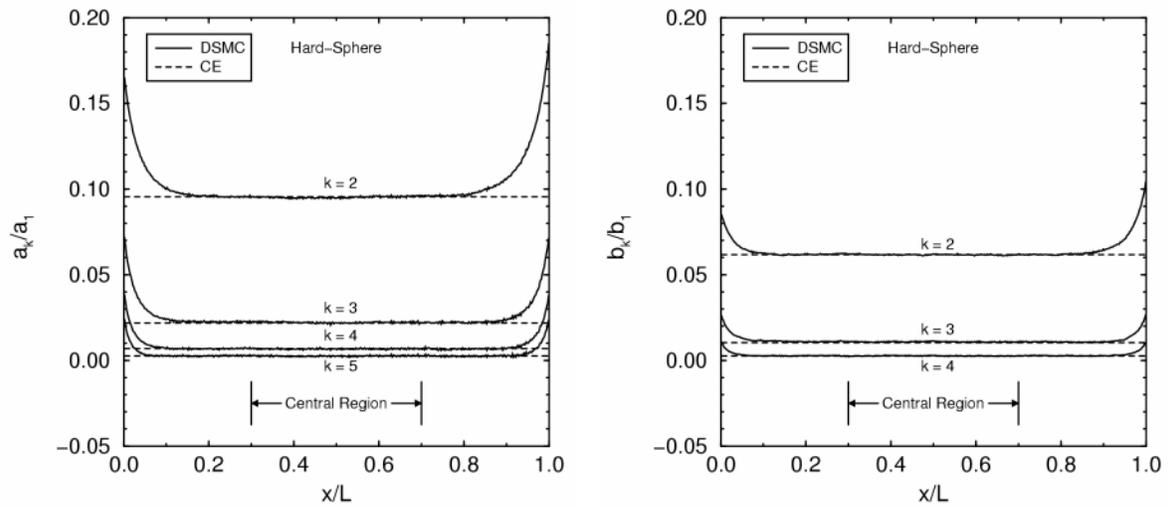

FIG. 7.

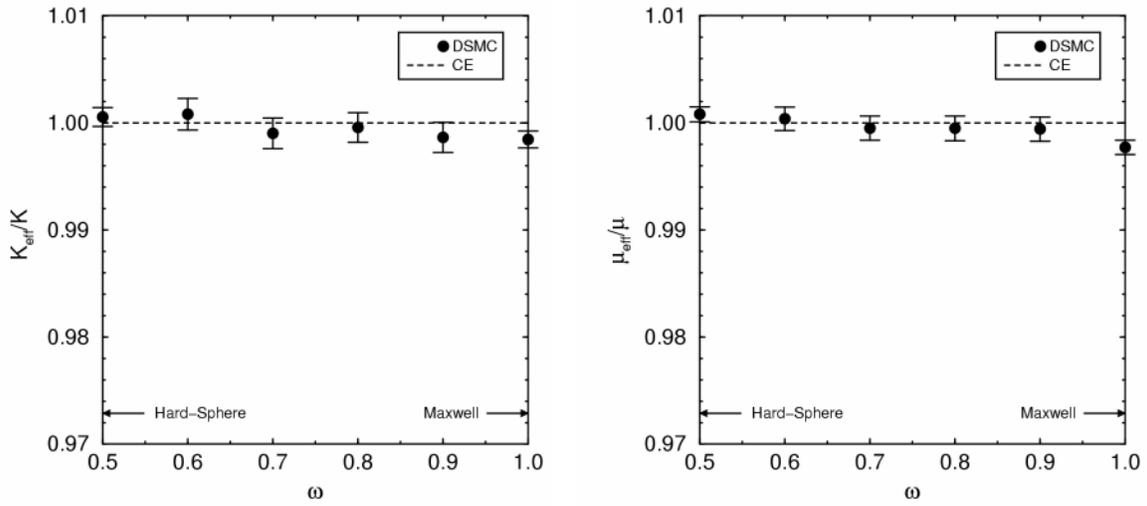

FIG. 8.

FIG. 9.

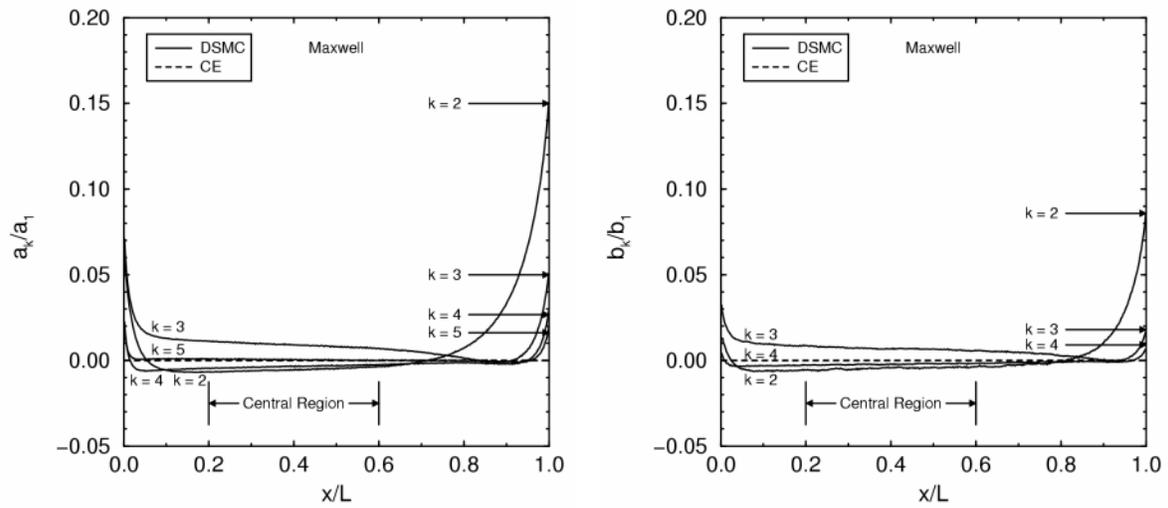

FIG. 10.

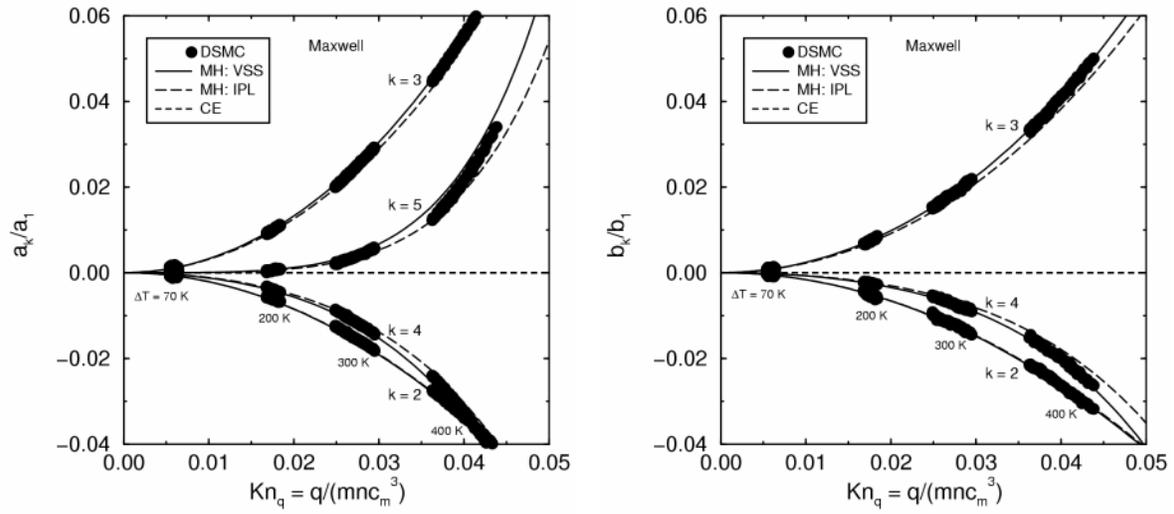

FIG. 11.

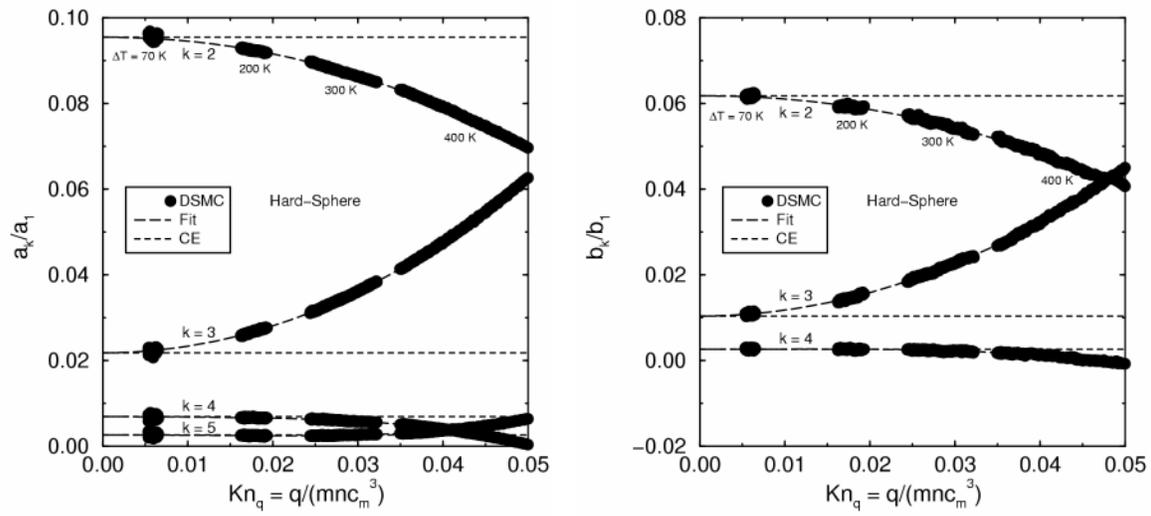

FIG. 12.

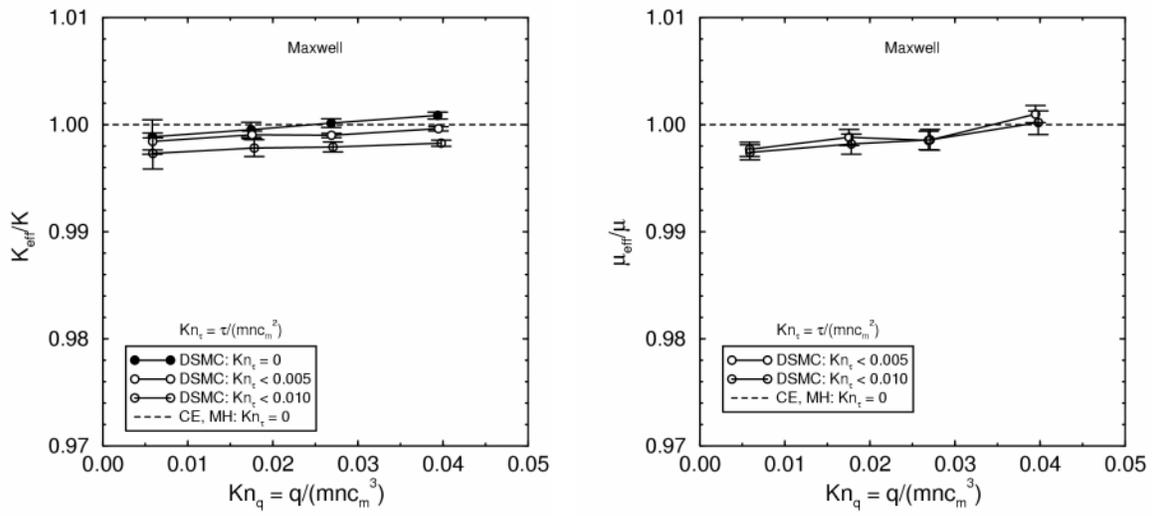

FIG. 13.

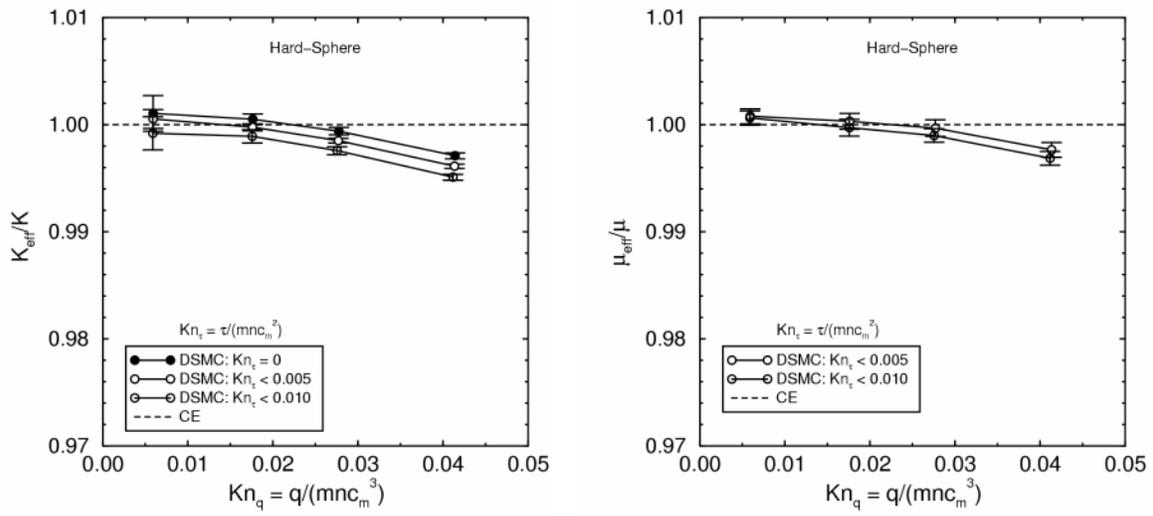

FIG. 14.

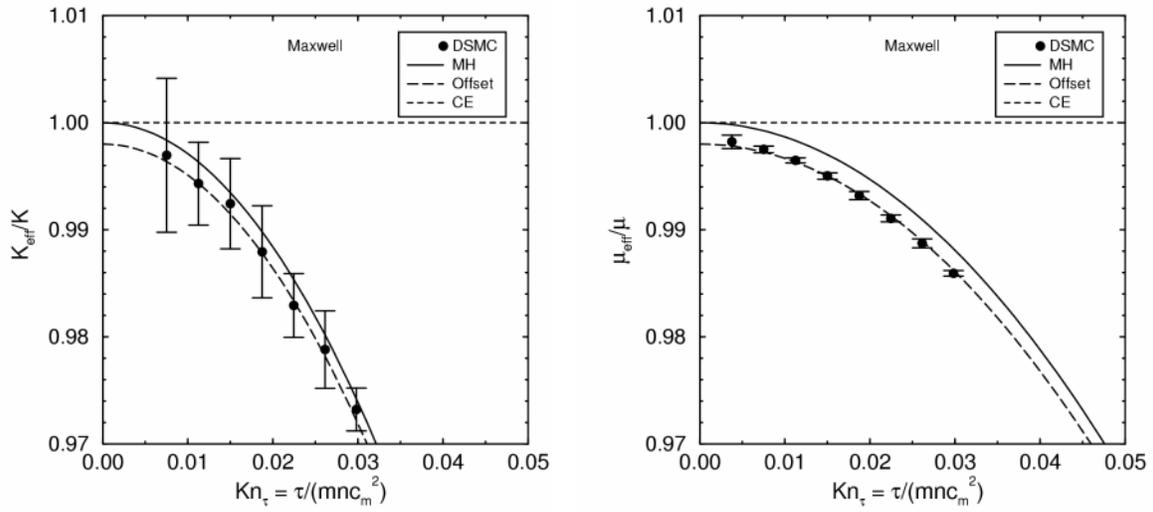

FIG. 15.

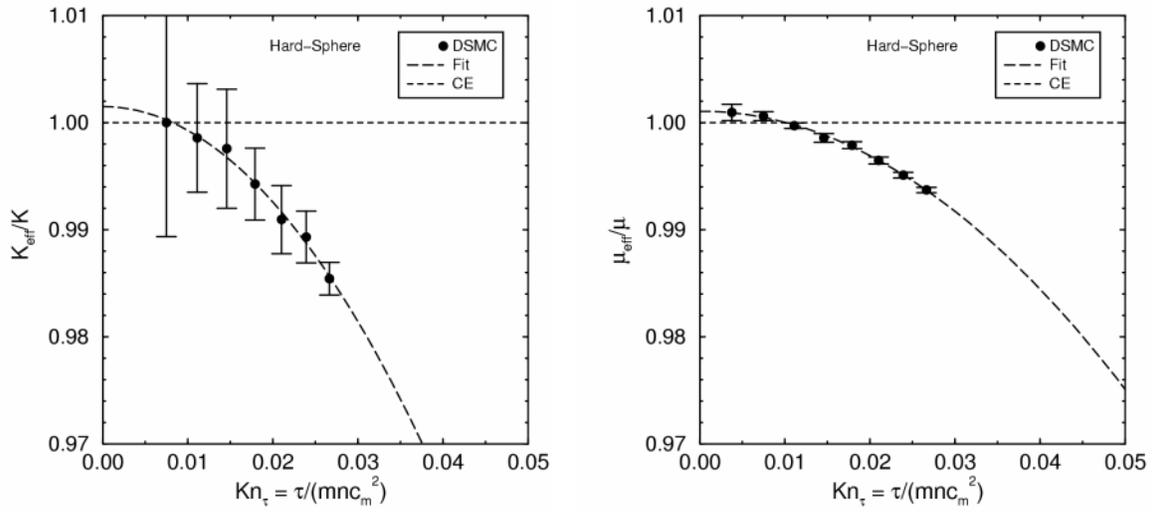